\documentclass[aip,cha,twocolumn,numerical author-year,reprint,floatfix]{revtex4}
\usepackage{amsfonts,amssymb,amsmath,wasysym,times}
\usepackage{color}
\usepackage{algorithm}
\usepackage{algpseudocode}
\usepackage{graphicx} 

\usepackage{url}
\usepackage[breaklinks]{hyperref}
\usepackage{breakurl}

\begin{document}

\begin{abstract}

Taking a pragmatic approach to the processes involved in the phenomena of collective opinion formation, we investigate two specific modifications to the co-evolving network voter model of opinion formation, studied by Holme and Newman \cite{holme_2006}. First, we replace the rewiring probability parameter by a distribution of probability of accepting or rejecting opinions between individuals, accounting for  the asymmetric influences in relationships among individuals in a social group. Second, we modify the rewiring step by a path-length-based preference for rewiring that reinforces local clustering. We have investigated the influences of these modifications on the outcomes of the simulations of this model. We found that varying the shape of the distribution of probability of accepting or rejecting opinions can lead to the emergence of two qualitatively  distinct final states, one having several isolated connected components each in internal consensus leading to the existence of diverse set of opinions and the other having one single dominant connected component with each node within it having the same opinion. Furthermore, and more importantly, we found that the  initial clustering in network can also induce similar transitions. Our investigation also brings forward that these transitions are governed by a weak and complex dependence on system size.  We found that the networks in the final states of the model have rich structural properties including the small world property for some parameter regimes.

\end{abstract}

\author{Nishant Malik} \email{nmalik@email.unc.edu}   \author{Peter J. Mucha} 
\affiliation {Department of Mathematics, CB 3250, University of North Carolina - Chapel Hill, NC 27599, USA}   

\title{Role of social environment and social clustering in spread of opinions in co-evolving networks}   
\maketitle

\section{Introduction} 

It has been widely reported in the media that online social networks like Facebook, Twitter, Blackberry messenger, etc.  played a key role in recent events in the world political sphere such as the Arab spring and London riots of 2011 \cite{scnews1,scnews2,scnews3,scnews4,pb1cs}. Meanwhile, there has also been increased interest in the quantitative and analytical analysis of the mechanisms and dynamics of the spread of social contagions such as rumors and opinions on complex networks \cite{stat-phy-sc-1,jp-on-1,pb1cs,centola1,dk1,jp-on-2,rmr-1,csnbook_2,csnbook_1,redner-3}. In such studies, individuals in the society are represented by nodes with edges indicating relationships between them, and  then techniques from statistical and nonlinear science are employed to analyze plausible models of the dynamics of spread of social contagions on a network \cite{holme_2006,bill_2011,vazquez_2008,kimura_2008,ww1,tc1,szm1,thilo1,thilo2,gleeson1,redner-1,redner-2}.

We propose a variation of the simplest coevolving network voter model of opinion formation, studied by Holme and Newman \cite{holme_2006}. In this model an edge is re-wired to connect two nodes having the same opinion, or the opinion of an individual is changed to agree with the the opinion of one of its neighbors based on a parameter, named the rewiring probability. We add two more simple mechanisms to this model, inspired by a  pragmatic approach to the modeling of asymmetric influences and tendencies to local clustering in the phenomenon of collective opinion formation in a social group so, that we can investigate a broader array of complex behaviors that can be induced by these modifications to the co-evolving voter model. For convenience of the exposition herein, we will refer to these additional mechanisms as : (1) \emph{Social Environment} and (2) \emph{Social Clustering}. Below we describe their meaning and significance in the processes of opinion formation.           

Acceptance and rejection of somebody else's opinions or choices by an individual depends on multitude of factors including the strength of relationship between the concerned individuals and the \emph{social environment} they live in. 
A prevailing \emph{social environment} (as defined for e.g.\ in \cite{se1}) not only alters relationships among individuals but can also affect their opinions on different issues in a fundamental way. A highly divisive society may be an outcome of inflexibilities in relationships that exist between individuals who resist accepting or sharing each others' opinions, choices or views. And these inflexibilities themselves could be due to the prevailing negative social environment. Other situations could involve positive social environment leading to flexible relationships among individuals hence leading to less resistance among individuals to the acceptance of each others' opinions, choices or views.  In modern times, media and advertising also play a significant role in altering the social environment and in constructing consent around certain opinions or choices \cite{nc1}. 

We propose to incorporate the effect of the social environment on the model of opinion formation on co-evolving networks by a distribution of probability of accepting or rejecting opinions between individuals. The distribution for social environment replaces the constant rewiring probability that has been used before in other studies on voter model with co-evolving networks \cite{holme_2006,bill_2011,vazquez_2008,kimura_2008}. Such description of the social environment becomes more plausible if we note the fact that relationships among individuals in a social group are inherently heterogeneous and asymmetric. For simplicity, we have assumed that the social environment is modulated by external social, economic or political factors   and its form remains the same over the temporal evolution of the model.   

Another important aspect that has not yet been sufficiently analyzed in the models of opinion formation on co-evolving networks has been the role of local clustering of edges in the network and other similar preferences for new links to be formed between nodes that are already near each other in the network. Indeed, in most models studied to date, the network distance has been considered to be  independent of the processes involved in the spread of opinions. In the present model we have attempted to explore the complex consequences of a simple introduction of such effects, by network distances and clustering in the network, with the processes of opinion formation. Specifically, we replace the random rewiring step of other models with a step that prefers  rewiring to nodes/actors who are both already closer within the network and who have higher probability of accepting new opinions. This way the  clustering of the evolving network in the model does not vanish in the large-network limit (as in other previous models). Clustering is a fundamental property of most network representations of social contexts,
 i.e., friends of friends have a higher likelihood (relative to the rest of the network) of also being friends \cite{csnbook_2,csnbook_1,strogatz_1998}. However, rewiring rules for co-evolving network models that do not reinforce clustering (as in, e.g., \cite{holme_2006,bill_2011}) can  randomize away any initial clustering, greatly simplifying the associated opinion dynamics. 

The explicit incorporation of model processes for \emph{social environment} and  \emph{social clustering} provides a simple simulation for the coupled effects of opinions with clustering and \emph{homophily}, the tendency of individuals to connect with individuals having similar characteristics \cite{homo1}.

\section{Description of the Model} 
Let  $G(N,E)$ be a network of $N$ nodes and $E$ edges with a predefined topology.  Let $\{ O_i \}$ represent  a set of $O$ number of opinions uniformly distributed over the $N$ nodes of $G(N,E)$ initially. Let $p_{ij}$ be the probability of some node $j$ accepting an opinion from node $i$. The distribution $P(p_{ij})$ describes the \emph{social environment}. If an edge exists between node $i$ and $j$ then we say $E_{ij}=1$. An edge connecting two nodes with different opinions is called a \emph{discordant edge} (i.e.,  where $E_{ij}=1$ but $O_i \neq O_j$). The total number of discordant edges in $G$ is represented by $E_{-}$ and $E=E_{+}+E_{-}$ where $E_{+}$ stands for harmonious edges (i.e., edges connecting nodes with the same opinion).       

\begin{algorithm}
\label{algo:algo1}
\begin{algorithmic}[1]
\State Generate a graph $G$ of given topology.
\State Generate a given distribution for $p_{ij}$ i.e. $P(p_{ij})$.  
\State Populate nodes with $O$ number of uniformly distributed opinions $\{ O_i \}$. 
\State   Calculate $E_{-}$. 
\While{$E_{-} \neq 0$}  {

\State Randomly choose a discordant edge $E_{ij}$.    
\State Generate a random number $\xi$ between $0$ and $1$ 
\If {$\xi<p_{ij}$} 
\State $O_j \gets O_i$ 
\State  Calculate $E_{-}$ 
\Else:
\State Remove the link between $i$ and $j$  i.e., set $E_{ij}=0$. 
\State Find a set ${\mathcal{N}'}$=$\{j\}_{j\neq i} \cap \{k\}$. 
\newline
\Comment{Where $\{j\}_{j\neq i}$ is a set containing all the nodes such that each element of it has $p_{ij}\geq \xi$ and $\{k\}$ contains all the nodes with shortest path from $i$ (excluding the nearest neighbours).} 

\If {${\mathcal{N}'}\neq \emptyset$} 

\State Connect $i$ randomly to any node $l \in {\mathcal{N}'}$
\State $O_l \gets O_i$

\Else:

\State Connect $i$ randomly to any node j s.t. $O_j=O_i$

\EndIf

\State Calculate $E_{-}$

\EndIf

}\EndWhile

\end{algorithmic}
\caption{A  hybrid voter model of opinion formation on a co-evolving network with clustering and distributed levels of influence.}
\end{algorithm}

 Different individuals have different probabilities of acceptance of others' opinions, which is here taken to be independent of the existence of a link between the individuals. Several factors ranging from socio-cultural affinity to the prevailing political and economic situation can influence these probabilities. To take these features into account we have used a distribution $P(p_{ij})$ for rewiring probabilities rather than a constant. Where $p_{ij}$ is the probability of $j$th node accepting the opinion of $i$th node.

We call $P(p_{ij})$  the \emph {social environment function}, accounting for the heterogeneous and asymmetric relationships among individuals. For the purposes of exploring a variety of settings, we have considered two different kinds of power laws for the \emph {social environment}. We set $P(p_{ij})=p_{ij}^\alpha$ to represent a \emph {flexible} social environment, i.e., individuals are able to accept others' opinion readily. Alternatively, we consider$P(p_{ij})=1-p_{ij}^\alpha$ to represent an  \emph {inflexible} social environment, i.e., individuals do not accept others' opinion readily and hence more churning happens in the society (see Fig.~{\ref{fig:fig_1}(b)). While there has been some empirical evidence to suggest that election results in multi-party democracies have power law distribution of votes among candidates from different parties  \cite{rncost1,els1,frd1}, however our use of a power law distribution in this specific context is driven only by its computational simplicity to simulate the qualitative kinds of social environment mentioned above \cite{accr1,mph1}. Other distributions such as exponential and extreme value distributions should also suffice to reproduce similar features. 

\begin{figure}
\centering
\includegraphics [height =1.9in]{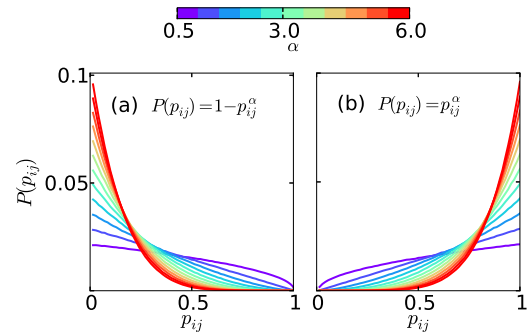}
\caption{(Color online) Different types of \emph{social environment function} $P(p_{ij})$, where $p_{ij}$ is the probability of $j$th node accepting the opinion of $i$th node.  (a) ``inflexible'', when we set $P(p_{ij})=1-p_{ij}^\alpha$, note that in this case more links will have lower probabilities of accepting opinions and (b) ``flexible'', when we set $P(p_{ij})=p_{ij}^\alpha$, note that in this case more links will have higher probabilities of accepting opinions.}
\label{fig:fig_1}
\end{figure}

Steps 13-16 in Algorithm 1 ensure that rewiring connections are mostly made according to \emph {social clustering} i.e., a node has higher probability of connecting to a person who is either a friend of a friend or, if no such connections are available, connecting to a person at the shortest possible distance identified in the network.
The set ${\mathcal{N}'}$ in the model (see Algorithm 1) consists of nodes/individuals who are close to some particular node $i$, both in terms of path length between them in the network and also they have higher probabilities of accepting the opinion of the node $i$. Hence, we call the nodes within the set ${N'}$ to be socially close to the node $i$. In case node $i$ is not able to find such individuals then it connects uniformly at random to somebody else holding the same opinion to avoid complete social isolation. Here, we aim to study the role of clustering of the network in altering the opinion space and network properties of the final end state. In so doing, our emphasis will be on transitions that occur in the network structure (notably, sizes and clustering of connected components) rather then just the space of opinions.} We will refer to the ratio of number of opinions to nodes i.e., $O/N$ as \emph{diversity}. We have fixed the average degree $\langle k \rangle=4$ and number of opinions $O=100$ for the simulations, if not mentioned otherwise. We have additionally investigated other numbers of opinions and average degree to confirm the robust nature of the qualitative properties described in this paper. The number of edges has been kept conserved throughout the dynamics; therefore at any time $t$, $E(t)= \langle k \rangle \frac{N}{2}$. Let the evolution of the system start at $t=t_\circ$ with $E_{-}(t_\circ)$ the initial number of discordant edges. The evolution of the system stops at the earliest such that $E_-(t_f)=0$, i.e., the final state of this model has no discordant edges left in the system. 

\begin{figure}
\centering
\includegraphics [height =2.75in]{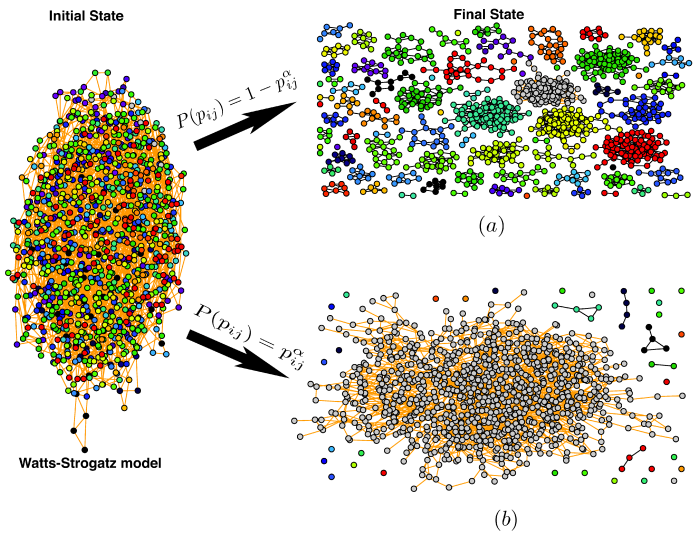}
\caption{(Color online)  A visual representation of the phenomena of formation of two distinct consensus states for two different social environment. Starting with an initial Watts-Strogatz network ($N=1000,\langle k \rangle=4$ and $\mathcal{C}=0.1$), we demonstrate the possibilities to end into two qualitatively different states. (a) $P(p_{ij})=1-p_{ij}^\alpha$ and $\alpha=6$ creates an ``inflexible'' social environment. We observe disintegration of the network into small connected components with each in internal consensus and having its own opinion, creating components with \emph{contrarian} positions i.e., \emph{segregated consensus} occurs in the network. (b) $P(p_{ij})=p_{ij}^\alpha$ and $\alpha=6$ creates a ``flexible'' social environment. We observe formation of dominant connected component in the final consensus state, having a size comparable to the initial network also, large number of opinions get extinct. We refer this kind of final state as the \emph{hegemonic consensus}.}
\label{fig:fig-vz1}
\end{figure}

There are several levels of plausible complexity for this model which could provide some new insights into the co-evolving dynamics of networks, but at the price of making it analytically harder to track. Indeed, even the limited analytical tractability of graph fission in a two-opinion co-evolving voter model presented in \cite{bill_2011} is undoubtedly aided by the rewiring rule considered there randomizing away all non-trivial clustering. In light of the complications introduced by the path-length influenced rewiring considered here, we have attempted to analyze this model computationally in a comprehensive way.

\subsection{Basic features of the model}

In this section we give a brief introduction to the basic features of this model. Firstly, we obtain two qualitatively distinct final states as we vary the social environment from flexible to inflexible. For a flexible social environment, if we set $P(p_{ij})=p_{ij}^\alpha$ with $\alpha=6.0$ then in the final state of the model, we observe formation of one single large connected component with each node having the same opinion and its size is comparable to the initial network. We call this kind of final state as the \emph{hegemonic consensus}, because of the emergence of one single hegemonic opinion.   
In the case of inflexible social environment, simulated by setting $P(p_{ij})=1-p_{ij}^\alpha$ with $\alpha=6.0$ we observe that initial network disintegrate into smaller isolated connected components where every node in each of these components hold the same opinion, i.e., each component is in the state of internal consensus. We will refer to this kind of final state as the \emph{segregated consensus}, as this feature is qualitatively similar to the \emph{segregation} of individuals in a society. A lattice based classical model of this social phenomena was given by Thomas Schelling \cite{tc1}, where he showed segregation of two groups of populations ('red' and 'white') who move over a check board following some simple rules. Several analytical and simulation results have been obtained following Schelling's model on networks as well as on co-evolving networks but not in context to the processes involved in collective opinion formation \cite{ems1,ems2,ems3}. Holme and Newman \cite{holme_2006},  observe some transitions qualitatively similar to that mentioned above by changing their constant rewiring probability parameter.

\begin{figure}
\centering
\includegraphics [width =\columnwidth]{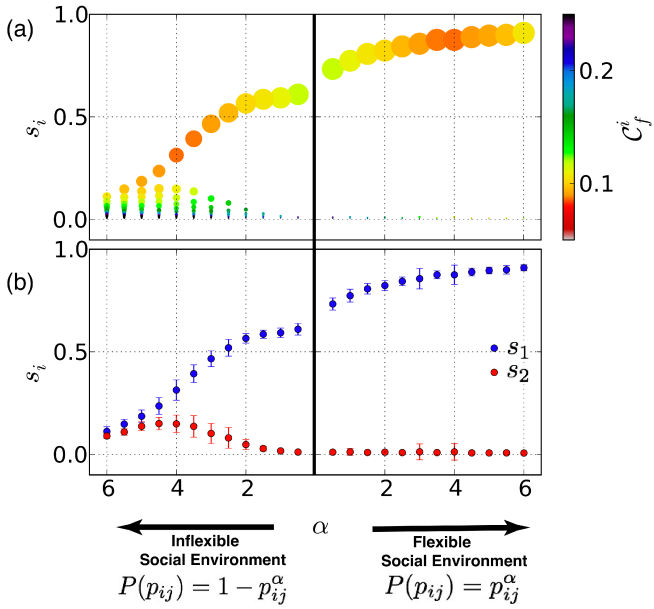}
\caption{(Color online) The effect of variation of social environment on a network of $N=1000$ nodes with $O=100$ opinions initially present. The starting network is an Erd\H{o}s-R\'{e}nyi random network  (i.e., clustering $\sim 1/N$). $\mathcal{C}_f^i$ is the clustering coefficient of the $i$th component in the final consensus state. (a) Size of a marker is proportional to the sizes of $s_i$, given by the fraction of nodes in the $i$th connected component. The thick bold line in the middle separates the two types of social environment. On the right we consider flexible social environment and we observe single large connected component with its size increasing with increasing $\alpha$. On the left consider inflexible social environment and we observe decreasing of the size of largest connected component with increasing $\alpha$, finally leading to its disintegration into several components of comparable sizes. (b) Sizes of two of the biggest connected components, $s_1$ is the size (as a fraction of the nodes in the network) of the largest connected component and $s_2$ is the size (as a fraction of the nodes in the network) of the second largest connected component. Simulations were carried over $100$ realizations of the network and opinion distribution. Sizes of the components are estimated as the mean over these realizations. Error bars give the standard deviation of these sizes over different realizations.}
\label{fig:figc1}
\end{figure}

A visualization for the above observation for $N=1000$ nodes with $O=100$ is shown in Fig.~\ref{fig:fig-vz1}. The drastic transition between the \emph{hegemonic consensus} and \emph{segregated consensus} in the final states of the systems seems to occur somewhere between the extreme flexible to inflexible social environment. Intuitively, it is perhaps not surprising that changing the distribution of the social environment induces a transition similar to that studied by Holme and Newman \cite{holme_2006}, insofar as the change in the distribution changes the overall average level of rewiring. Nevertheless, a priori we have no reason to expect that change in the form of the distribution of probabilities of accepting or rejecting of others' opinions should have similar effects as the changes to the single rewiring probability parameter employed by Holme and Newman \cite{holme_2006}. Also, the detailed structural properties of  the  network in the \emph{hegemonic consensus} and \emph{segregated consensus} in the final state are expected to be much richer  as shown and discussed below in some detail. In Fig.~\ref{fig:figc1} we observe the effect of varying the social environment, where $s_{i}$ is the size (fraction of nodes) of the $i$th component in the final consensus state with $i=1$ being the largest component. A further analysis of the phase transition involved in emergence of these two distinct states in this system has been attempted in detail in the following section, as one of the two central themes of this paper. 

\begin{figure}
\centering
\includegraphics [width =\columnwidth]{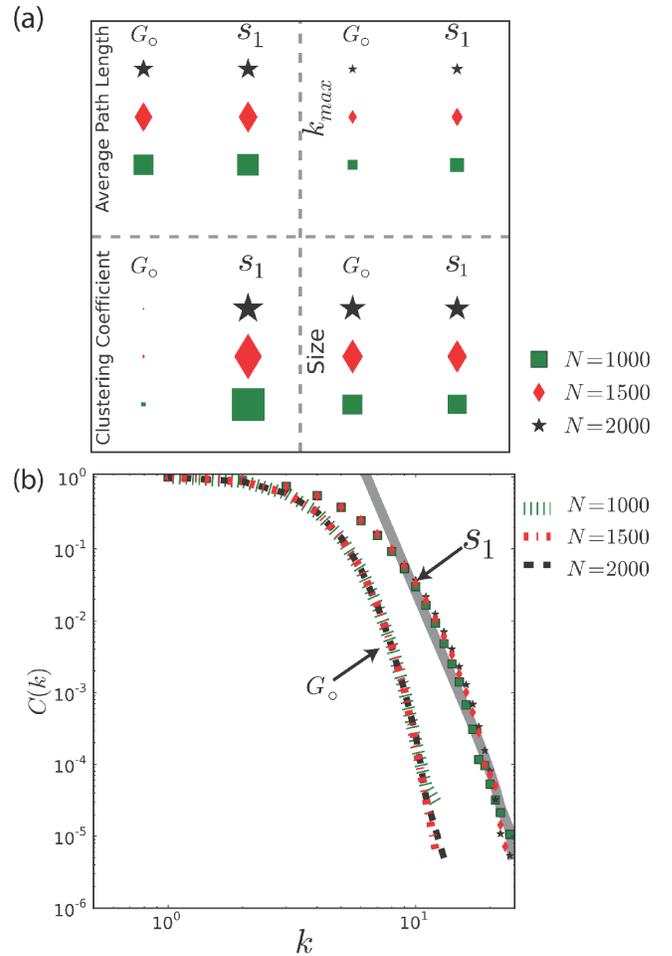}
\caption{(Color online) Properties of the largest connected component in \emph{hegemonic consensus} :  The final state reached for the flexible environment $P(p_{ij})=p_{ij}$ with $\alpha=6.0$ when starting with an Erd\H{o}s-R\'{e}nyi network. (a) Different markers represent different initial network sizes (see the legend). The initial network is $G_0$, its initial clustering is close to zero and $s_1$ is the size (as a fraction of the nodes in the network) of the largest connected component in the final state. We observe that $s_1$ has a significantly higher clustering coefficient ($0.2$). Whereas it has comparable small path length to the initial  Erd\H{o}s-R\'{e}nyi network $G_0$, implying that $s_1$ has small world features. Also, $s_1$ has in general higher $k_\mathrm{max}$ (maximum degree) and its size is comparable to $G_0$. (b) Shows the the cumulative degree distribution $C(k)$ of the initial network $G_0$ (dashed lines) and $s_1$ (markers). $s_1$ does have nodes with higher degrees. In its tail, the cumulative degree distribution of $s_1$ appears to approximately follow a power law as shown by solid grey line of exponent $-8$.}
\label{fig:figplx1}
\end{figure} 
\begin{figure}
\centering
\includegraphics [width =0.95\columnwidth]{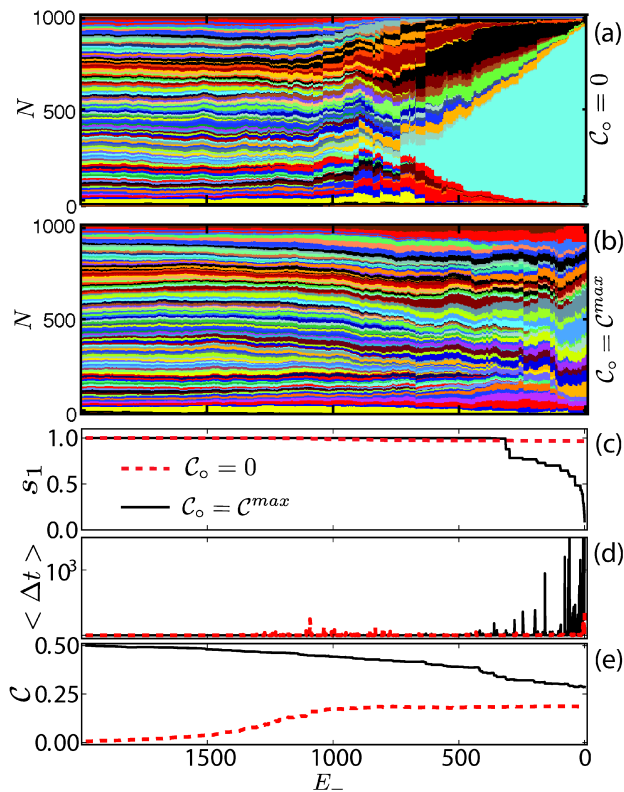}
\caption{(Color online) The evolution of different variables in the system with decreasing number  of discordant edges. Each variable is plotted at time step when that number of discordant edges, $E_{-}$  was present for the last time in the system. The black line and panel (b) corresponds to simulations starting at the highest possible clustering coefficient $\mathcal{C}^{\mathrm{max}}$, whereas red dotted line and panel (a) corresponds to simulations starting at the negligible clustering coefficient (random network). In (a)  and (b) each color corresponds to one of the opinions, width of each color gives the number nodes occupying that opinion. In (a) note the wide width of cyan color at the end, this represents the formation of \emph{hegemonic consensus} (one large connected component of size comparable to initial network and with each node being at consensus with every other node.) In (b) we does not observe this transition, only difference in this simulation is the large initial clustering coefficient. (c) $s_1$ is the size of largest connected component. Observe the abrupt drop in $s_1$ in the case of the black line, this indicates transition to the disintegration of network into smaller components i.e., \emph{segregated consensus}. In contrast we do not observe any such transition for the red dotted line which corresponds to the formation of \emph{hegemonic consensus}. (d) $\langle \Delta t \rangle$ is the average number of iterations the system takes to the removal of single discordant edge. It shows a substantial increase for the black curve at the end. (e) $\mathcal{C}$ is corresponding evolution of the clustering coefficient.}
\label{fig:fig_ts}
\end{figure}

The giant consensus community occurring in the Holme and Newman model \cite{holme_2006} would appear to be structurally similar to networks obtained under a configuration model with the observed final state degree distribution. In contrast, as observed in Fig.~\ref{fig:figplx1}(a) the largest connected component in the \emph{hegemonic consensus} has small world properties (average path lengths comparable to random network and high clustering coefficients)  and it also consists of nodes with higher number of connections as apparent from the change in cumulative degree distribution as shown in Fig.~\ref{fig:figplx1}(b). These features are closer to organized political or religious movements, which usually have a hierarchy of leadership and high clustering, thus we have pointedly not referred to this structure as a \emph{mob}, because of the observed hierarchy of connectivity involved here. We have not observed variation in \emph{diversity} $O/N$ to bring about any significant change to the above discussed basic properties, while varying the values of $O$ from $2$ to $100$. 

\begin{figure}
\centering
\includegraphics [width=0.85\columnwidth]{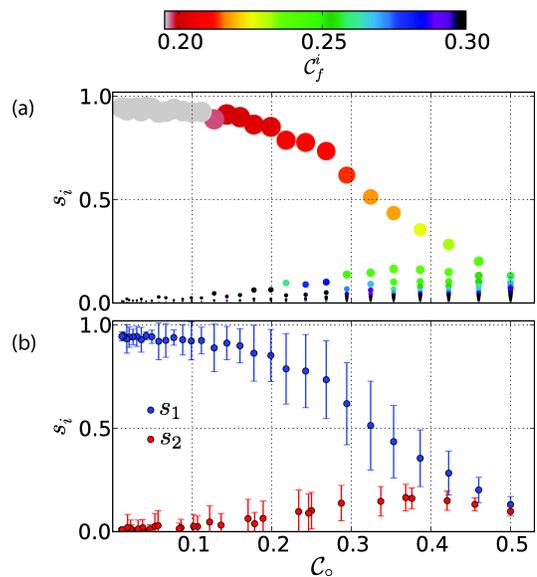}
\caption{(Color online) The effect of variation of initial clustering $\mathcal{C}_\circ$ on a network of N = 1000 nodes with O = 100 opinions initially present when social environment is flexible with $P(p_{ij})=p_{ij}^\alpha$ with $\alpha=6$. Increasing of initial clustering $\mathcal{C}_\circ$, leading to the transitions i.e., disintegration of network in consensus state into smaller connected component (\emph{segregated consensus}) contrary to the expected \emph{hegemonic consensus} for initially unclustered networks in flexible social environment. (a) Size of a marker is proportional to the size of $i$th connected component in the final consensus state. Observe the disintegration of the network into several connected components for higher values of initial clustering $\mathcal{C}_\circ$. (b) $s_1$ is the size of largest connected component and $s_2$ is the size of the second largest connected component. Simulations were carried over $100$ realizations of the network and opinion distribution. Sizes of the components are estimated as the mean over these realizations. Error bars gives the standard deviation of these sizes over different realizations. $\mathcal{C}_f^i$ is the clustering coefficient of the $i$th component in the final consensus state. Observe the higher values for connected components after the disintegration into smaller components. }
\label{fig:fig-ts-cls}
\end{figure}

Another crucial aspect to consider in this model is the role of initial network topology in transitions between \emph{hegemonic consensus} and  \emph{segregated consensus} as the two distinct final states. Does the variation of the initial clustering coefficient change the final state? This question have not been considered in the previous studies of voter model on co-evolving networks, as the previously introduced models have not treated clustering as 
 a consequence of those models, even though clustering  is one of the essential characteristics of social networks \cite{csnbook_2,csnbook_1,strogatz_1998}. In the model considered here, the formation of a \emph{hegemonic consensus} state apparently does not take place in networks with high initial clustering coefficient. To understand this feature we investigate the evolution of the clustering in this model. 
 
\begin{figure*}
\centering
\includegraphics[scale =0.3]{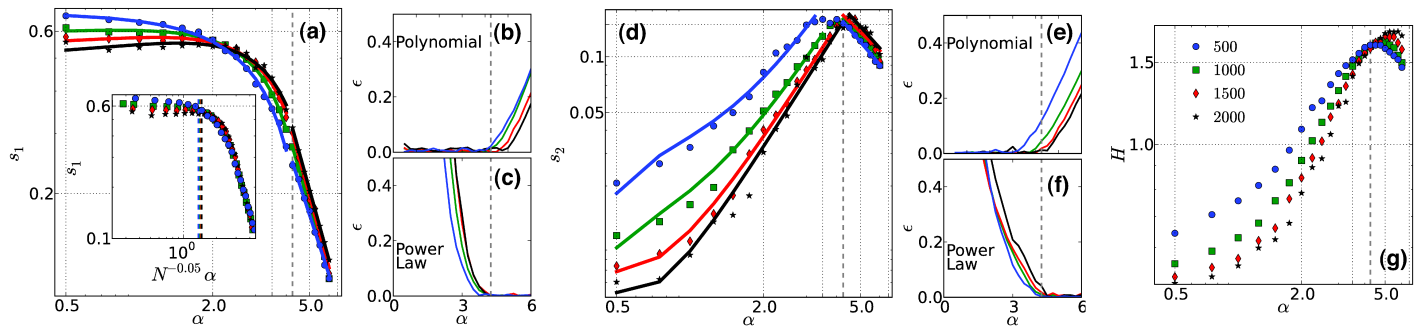}
\caption{ (Color online) Variation of $s_1$ (size of largest connected component) and $s_2$ (size of second largest connected component) with $\alpha$. Here social environment is set to be \emph{inflexible}, i.e., $P(p_{ij})=1-p_{ij}^\alpha$. Different shapes and colors of the  markers represent networks of different sizes (see legend in (g)).  In (a) we observe multiple transitions in $s_1$,  first happens at $\alpha= N^{0.05}$ where all data point collapse onto the same curve (see inset). A second transition is observed at $\alpha= 4.25$ (dashed grey vertical line), where a best fit to the data changes from a polynomial to power law (see (b) and (c), where $\epsilon$ gives the error between the between fitted function and the data points.). This second transition also appears in an even more visually apparent form in (d),
observe the abrupt decreasing of $s_2$ after $\alpha= 4.25$ (dashed grey vertical line). Again best fit to the data changes from a polynomial to power law (see (e) and (f)). In the figure (g) we plot Shanon's entropy $H$ of the $10$ largest  connected components versus $\alpha$, in this figure too, we observe $H$ tends to saturate at $\alpha= 4.25$ (dashed grey vertical line) and start to decrease after linear increase.}
\label{fig:alp_tr}
\end{figure*}

Let the \emph{clustering coefficient} of the network be represented by the symbol $\mathcal{C}$,  defined as three times the ratio of the number of loops of length three in the network to the number of connected triples of nodes, also known as \emph{transitivity} \cite{sndoro1}. Symbols $\mathcal{C}_\circ$ and $\mathcal{C}_f$ are used here for the initial clustering at start of the simulation and final clustering at the end of the simulation, respectively. In the Watts-Strogatz  model, for example the maximum possible initial clustering $\mathcal{C}^{\mathrm{max}}$ corresponding to the ring topology, is $\mathcal{C}^{\mathrm{max}} = \frac{3(\langle k \rangle-2)}{4(\langle k \rangle-1)}$. Therefore, with $\langle k\rangle =4$, we would have $\mathcal{C}^{\mathrm{max}} = 0.5$ (see e.g. \cite{nw1}). The $C^\mathrm{max}$ value is also the upper limit for $\mathcal{C}_f$. In Fig.~\ref{fig:fig_ts} we plot the evolution of different variables of the model from a single simulation as discordant edges are removed. The social environment was set to be flexible  $P(p_{ij})=p_{ij}^\alpha$ with $\alpha=6.0$, i.e., the parametric regime where we expect formation of a \emph{hegemonic consensus} state for initially unclustered networks. The size of the initial network was $N=1000$. When we set $\mathcal{C}_\circ=0$, the opinion space does undergo a transition as expected and we see one opinion dominating (see Fig.~\ref{fig:fig_ts} (a)). Also to be noted at the same time there is no transition in the size of the largest connected component (see red dotted line in Fig.~\ref{fig:fig_ts} (c)). For the black curve in Fig.~\ref{fig:fig_ts} we have set $\mathcal{C}_\circ=\mathcal{C}^{\mathrm{max}}$ and we observe a counter intuitive and unexpected transition viz., that the largest connected component starts to disintegrate and become smaller in size (see Fig.~\ref{fig:fig_ts}(c)) and also in opinion space we do not observe emergence of a single dominant opinion (see Fig.~\ref{fig:fig_ts}(b)). We also observe in the lowest panel of Fig.~\ref{fig:fig_ts} that $\mathcal{C}$ saturates to  $\mathcal{C}_f$ before reaching the consensus. This is a special feature of this model and provides this opportunity to study the evolution of a clustered network topology with opinion formation. For the case  $P(p_{ij})=p_{ij}^\alpha$ with $\alpha=6$, $\mathcal{C}_f$  appears to be well approximated by a linear function of $\mathcal{C}_\circ$. We also see in the panel (d) of Fig.~\ref{fig:fig_ts} that right before the consensus states emerges, the system start to slow down. That is, more iterations are required to decrease the number of discordant edges, possibly indicative of some form of critical slowing of the system as \emph{segregation} is reached. This feature is not so apparent in case of red dotted curve, implying that processes involved in formation of \emph{hegemonic consensus} do not involve critical slowing of the system. In Fig.~\ref{fig:fig-ts-cls} we show the disintegration of the network into smaller components as we increase the initial clustering coefficient from $0$ to $\mathcal{C}^{\mathrm{max}}$. The above discussion only briefly illustrates some of the features in the evolution of clustering in the model. Below, we would present a systematic analysis of this transition.

\section{Phase transitions} 
\subsection{Role of social environment in transitions}

As discussed above this model shows transition to two distinct final states, for \emph{flexible} social environment we have observed that as $\alpha$ is increased, the largest connected component's size approaches that of the whole network,  $s_1 \to  1$ (see Fig \ref{fig:figc1})  and each node within the component hold the same opinion, and for the  \emph{inflexible} social environment case we have disintegration of the network into several smaller sized connected components, where nodes within each of the components holds the same opinion. As we move from inflexible to flexible social environment fewer and fewer initial opinions survive, with the most extreme case being where only one dominant opinion survives with formation of a \emph{hegemonic consensus}. Here we will attempt to infer from numerical simulations whether these transitions have a finite size effect \cite{fts1}. The complexities involved in this model makes analytical analysis hard but it is possible to obtain a variety of details using numerical simulations.

From Fig.~\ref{fig:figc1} we observe that somewhere when the parameters of the model are in the \emph{inflexible} social environment regime there is emergence of smaller sized connected components. Hence, we will examine transition within parameter setting of \emph{inflexible} social environment i.e., $P(p_{ij}) =1-p_{ij}^\alpha$. The initial network for all the simulation below is Erd\H{o}s-R\'{e}nyi random network. In Fig.~\ref{fig:alp_tr}(a) we observe rather multiple transitions in the system when $\alpha$ is varied from $0.5$ to $6.0$ for the inflexible social environment. The first transition is visible in the size of $s_1$, where a weak dependence on the size of the system seems to emerge (see inset Fig.~\ref{fig:alp_tr}(a)). All the curves with different system sizes collapse onto  one single curve when a small factor $N^{-0.05}$ is multiplied to $\alpha$ that is, it appears that this transition point has dependence on the size of the system and it would change as $\alpha=N^{0.05}$ (see vertical lines in the inset of \ref{fig:alp_tr}(a)) and this transition point would move to infinity in the thermodynamic limit.

A second transition occurs at $\alpha=4.25$ where the best fit to the data points turns from a polynomial fit to power law fit (see Fig.~\ref{fig:alp_tr}(b-c) and (e-f)). For fitting functions we have used a least squares routine provided in SciPy's  optimize package, which uses MINPACK's lmdif and lmder algorithms \cite{scipy-1}. This transition is more apparent in Fig.~\ref{fig:alp_tr}(d) for the size of the second largest connected component i.e., $s_2$. In Fig.~\ref{fig:alp_tr}(g) we have plotted the Shannon entropy over the sizes of the $10$ largest components with $H=\sum_{i=1}^{i=10}s_i\ln(s_i)$. Considering only $10$ largest components for this calculation is a reasonable approximation to the total Shannon entropy of the size distribution in most cases, given the rapid decrease in the tail of the size distribution. In this figure as well the transition at $\alpha=4.25$ is visibly very much apparent as $H$ tends to saturate and start to decrease after linear increase. The polynomial fit in Fig.~\ref{fig:alp_tr} (a)) has the following form :  \begin{align} 
s_1  = & a\alpha^{2}+b\alpha+c  & \text{if}  ~~ \alpha < 4.25   \nonumber \\
s_1  \thicksim & f(N)\alpha^{-2.4 \pm 0.02} & \text{if}  ~~ \alpha \geq 4.25 
\end{align} where $a  \thickapprox -0.02 9$, $b \thickapprox N^{0.052 \pm 0.001}-1.4$ and $c \thickapprox N^{-0.36}\log(N)$  and $f(N)$ is function dependent on $N$. A similar analysis for $s_2$ also yields a polynomial fit : 
\begin{align}
s_2  = & a\alpha^{-2.1}+b\alpha^{2.1}+c   &\text{if}  ~~ \alpha < 4.25  \nonumber \\
s_2  \thicksim & f(N)\alpha^{1.42\pm 0.12}  &\text{if} ~~ \alpha \geq 4.25
\end{align} where $a \thicksim N^{0.0027}-1.02$, $b \thickapprox -2.68 \time 10^{-6} N-1.54$ and $c \thicksim N^{1.75}$ and again $f(N)$ is function dependent on $N$. This analysis brings out a highly complex dependence of $s_1$ and $s_2$ on system size for the transition occurring at $\alpha = 4.25$. But as indicated by the error to polynomial fit and power law fits in Fig.~\ref{fig:alp_tr}(b-c) and (e-f), a polynomial fit becomes systematically less erroneous as $N$ is increased. Which means for large $N$ these multiple transitions might coalesce into one single continuous transition.   

\subsection{Role of network structure in transitions}  
\label{sec:sec_C}

Social networks are generally known to have higher clustering ~\cite{snet1}. The initial definition of global and local clustering was in the context of social ties ~\cite{csnbook_2,csnbook_1,strogatz_1998,trans_1971}. In previously studied coevolving voter models with random rewiring the clustering tends to decay away to that of independently distributed edges ($\sim 1/N$) as the system evolves with time ~\cite{holme_2006,kimura_2008,vazquez_2008,bill_2011}. Whereas in the present model we observe that a net critical value is sustained throughout its evolution and never dropping to near zero (see Fig.~{\ref{fig:fig_ts}). 

\begin{figure}
\centering
\includegraphics [width=0.85\columnwidth]{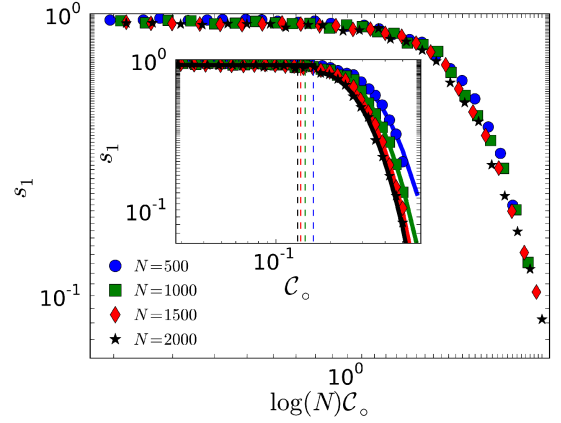}
\caption{Variation in the size of largest connected component $s_1$ with the initial clustering coefficient. The social environment was fixed to be \emph{flexible} i.e., $P(p_{ij})=p_{ij}^\alpha$ with $\alpha=6$. When $\log(N)$ is multiplied to $\mathcal{C}_\circ$ the data for different system sizes collapses onto same curve. The inset curve shows the fits and the vertical lines are $\frac{1}{\log(N)}$, indicating the transition points.} 
\label{fig:clus1tr}
\end{figure}

Such a model provides an opportunity to explore the influence of variation in the clustering coefficient on transitions between the formation of a \emph{hegemonic consensus} and \emph{segregated consensus}. We are here mainly interested in knowing whether $\mathcal{C}_\circ$, the initial clustering, can affect the formation of the \emph{hegemonic consensus}. We know from the discussion above that if we set $P(p_{ij})=p_{ij}^\alpha$ and $\alpha=6$ (flexible social environment), we will get the \emph{hegemonic consensus} to be the final  state, where the size of the largest connected component $s_1 \sim 1$ in the consensus state for an initial random network of independently distributed edges (or network with negligible clustering coefficient). After setting $P(p_{ij})=p_{ij}^\alpha$ with $\alpha=6.0$ we vary the initial clustering $\mathcal{C}_\circ$ of the system, employing a Watts-Strogatz model for the initial network. We observe in inset of Fig.~\ref{fig:clus1tr} that with increasing initial clustering, the largest connected component does tend to disintegrate into smaller size. For higher $\mathcal{C}_\circ$, rather then having only one dominant connected component of size $s_1 \sim 1$, we get smaller sized connected components, i.e., \emph{segregated consensus} occurs in place of \emph{hegemonic consensus}. So, even in the case of a highly flexible social environment i.e., $P(p_{ij})=p_{ij}^\alpha$ and $\alpha=6$, we can still get disintegration and no single dominant opinion, if the initial clustering of the network is high enough.  

To get an estimate on the values $\mathcal{C}_\circ$, where we could start observing the disintegration in the consensus state we further analyze the results obtained in Fig.~\ref{fig:clus1tr}.  We observe that if we multiply a factor $\log(N)$ to the $\mathcal{C}_\circ$ then all the data collapses onto one curve (see Fig.~\ref{fig:clus1tr}) implying that transition seems to be occurring at $\mathcal{C}_\circ= \frac{1}{\log(N)}$. If we plot the transition points  $\frac{1}{\log(N)}$ as done in the inset of Fig.~\ref{fig:clus1tr} by means of vertical lines, we do observe spontaneous drop off in the values of $s_1$ around these transitions. The form of the function that can be fitted to the data in Fig.~\ref{fig:clus1tr} is as follows : 

$$
s_1 \sim  \left\{ \begin{array}{rl}
  
  1 ~~~~~~~~~~~~~~~~~~						   & \mbox{if $\mathcal{C}_\circ \leq \frac{1}{\log(N)}$}  \\*
  a \mathcal{C}_\circ^\alpha\exp(-\lambda\mathcal{C}_\circ)  & \mbox{if $\mathcal{C}_\circ > \frac{1}{\log(N)}$} 
  
       \end{array} \right.
$$ where, $\lambda \sim N^{-0.37\pm 0.018}$, $a \sim N^{-0.95\pm0.07}$ and $\alpha~\sim N^{-0.13\pm 0.012}$. Though the above functional form might has a complex dependence on the system sizes, the critical values  $\mathcal{C}_\circ$ are clearly varying as  $\frac{1}{\log(N)}$ (see vertical lines in inset of Fig.~\ref{fig:clus1tr}). Hence, this transition would exist in a finite network and the critical value of $\mathcal{C}_\circ$ would become zero in the thermodynamic limit.

\begin{figure}
\centering
\includegraphics [width=\columnwidth]{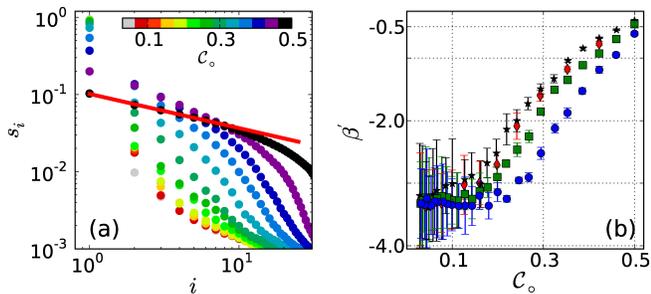}
\caption{(Color online) The sizes of different connected components in the consensus state for network of $N=1500$ nodes. (a) Sizes of connected components v.\ their ordered indices.  The largest component has index $1$ and indices are arranged in decreasing order of sizes of the components on abscissa. As initial clustering of the network $\mathcal{C}_\circ$ (color bar) is increased, there is emergence of smaller components of comparable sizes. (b) $\beta'$  are the values of the exponents of the slopes fitted to the sizes of components in the final consensus state v.\ indices at each value of $\mathcal{C}_\circ$ (thick red line in (a) is an example for the same for $\mathcal{C}_\circ=0.5$). In (b) observe the decrease in the slope and error bars for higher initial clusterings, indicating the formation of several components of comparable sizes.}
\label{fig:clus2sl}
\end{figure}

A further analysis of the connected components formed in \emph{segregated consensus}  shows that their sizes are power law distributed.  In Fig.~\ref{fig:clus2sl}(b) we have plotted the slope of the line fitted to the sizes of connected components and  in Fig.~\ref{fig:clus2sl}(a) there is an illustration of the same for $N=1500$ nodes.  As we increase $\mathcal{C}_\circ$ not only the slope becomes smaller, but also the error bar to the fit is reducing indicating that sizes of the connected components are becoming comparable as $\mathcal{C}_\circ$ is increased, i.e., similar sized contrarian social groups or \emph{cults} are formed. We also note from  Fig.~\ref{fig:fig-ts-cls} that these similar sized components generally have very high clustering.  
  
\section{Conclusions} 

We considered a model for the opinion formation on coevolving networks with two additional attributes: one is the social environment, which is modeled by a distribution of susceptibilities to opinion change, and the second one is a path-length-based preference for rewiring that reinforce social clustering. The social clustering component intrinsically links the topological evolution of the network with the processes involved in collective opinion formation and vice versa. We observed that two qualitatively distinct final states can emerge in this model, in one where we have formation of \emph{hegemonic consensus}, a dominating large connected component with each node having the same opinion. Importantly, this dominating large connected component also maintains nontrivial local clustering. Such clustering contrasts with the properties of previously studied models, as random rewiring in them leads to non-clustered random networks as the final consensus state. 

The other outcome that emerges under the parameter settings of inflexible social environments is the disintegration of the network and formation of small isolated components consisting of nodes holding the same opinion. It is a feature qualitatively similar to the \emph{segregation} of individuals in a society due to the internal conflicts and frustrations leading to formation of dysfunctional social networks. Hence, we have named this final state as \emph{segregated consensus}.     

A fundamentally key aspect we studied was the role of clustering in the network in the process of opinion formation on co-evolving networks using the features of this model, where the clustering of the network is continually re-inforced by the preference to rewire to nodes at smaller path length in this model. We observed that if the initial network has clustering above a critical value, then even in a flexible social environment we get \emph{segregated consensus} as the final state. This is contrary  to what happens when we start with a network having negligible clustering (random network). Injection of this additional attribute to the model makes the dynamics of this system richer and more interesting but at the price of making any analytical study much more difficult than for other models, such as discussed in \cite{bill_2011,vazquez_2008,kimura_2008,ww1,tc1,szm1,thilo1,thilo2}.

One can observe similar features in the process of opinion formation in society, for example \emph{hegemonic consensus} can be analogous to situations in the states with multi-party democratic elections, where one party wins  by a landslide. In contrast, some hung elections may
 be similar to a \emph{segregated consensus}  \cite{mdemo-1}. A similar situation can also occur when choices are made on a product  among the many available brands, with monopoly of one brand over the product being the \emph{hegemonic consensus}  and  \emph{segregated consensus} being when there is more even competition over a product between different brands \cite{mdemo-2}. 

Further analysis of the transitions in numerical simulations of different sizes indicated complex and weak dependence on the system size. In particular, it is possible that the multiple transitions induced by variations in social environment might coalesce into one single continuous transition for large systems. Meanwhile, the transition induced by clustering in the initial network only exists for a finite system. Importantly, because this latter transition occurs for initial clustering $\sim 1/\log(N)$ (cf.\ independently distributed edges yield clustering $\sim 1/N$), we note that one should be careful making any claims about the applicability of coevolving network models that lack reinforcement of clustering to real-world network situations that have non-trivial transitivity.

\acknowledgments The authors thank Mason Porter and Feng Shi for helpful suggestions and comments. The project described was supported by Award Number R21GM099493 from the National Institute of General Medical Sciences. The content is solely the responsibility of the authors and does not necessarily represent the official views of the National Institute of General Medical Sciences or the National Institutes of Health.


\begin{thebibliography}{46}
\expandafter\ifx\csname natexlab\endcsname\relax\def\natexlab#1{#1}\fi
\expandafter\ifx\csname bibnamefont\endcsname\relax
  \def\bibnamefont#1{#1}\fi
\expandafter\ifx\csname bibfnamefont\endcsname\relax
  \def\bibfnamefont#1{#1}\fi
\expandafter\ifx\csname citenamefont\endcsname\relax
  \def\citenamefont#1{#1}\fi
\expandafter\ifx\csname url\endcsname\relax
  \def\url#1{\texttt{#1}}\fi
\expandafter\ifx\csname urlprefix\endcsname\relax\def\urlprefix{URL }\fi
\providecommand{\bibinfo}[2]{#2}
\providecommand{\eprint}[2][]{\url{#2}}

\bibitem[{\citenamefont{Holme and Newman}(2006)}]{holme_2006}
\bibinfo{author}{\bibfnamefont{P.}~\bibnamefont{Holme}} \bibnamefont{and}
  \bibinfo{author}{\bibfnamefont{M.}~\bibnamefont{Newman}},
  \bibinfo{journal}{Phys. Rev. E} \textbf{\bibinfo{volume}{74}},
  \bibinfo{pages}{056108} (\bibinfo{year}{2006}).

\bibitem[{\citenamefont{Srinivasan}(August 11, 2011)}]{scnews1}
\bibinfo{author}{\bibfnamefont{R.}~\bibnamefont{Srinivasan}},
  \emph{\bibinfo{title}{London, Egypt and the nature of social media}}
  (\bibinfo{publisher}{The Washington Post}, \bibinfo{year}{August 11, 2011}),
  \urlprefix \href{http://articles.washingtonpost.com/2011-08-11/national/35271787_1_social-media-social-networks-mobile-communication}{\tt{http://articles.washingtonpost.com/2011-08\linebreak-11/national/35271787\_1\_social-media-social-\linebreak networks-mobile-communication}}.

\bibitem[{\citenamefont{Ball and Lewis}(August 24, 2011)}]{scnews2}
\bibinfo{author}{\bibfnamefont{J.}~\bibnamefont{Ball}} \bibnamefont{and}
  \bibinfo{author}{\bibfnamefont{P.}~\bibnamefont{Lewis}},
  \emph{\bibinfo{title}{Riots database of 2.5m tweets reveals complex picture
  of interaction}} (\bibinfo{publisher}{The Guardian}, \bibinfo{year}{August
  24, 2011}), \urlprefix
  \href{http://www.guardian.co.uk/uk/2011/aug/24/riots-database-twitter-interaction}{\tt{http://www.guardian.co.uk/uk/2011/aug/24/\linebreak riots-database-twitter-interaction}}.

\bibitem[{\citenamefont{Huang}(June 6, 2011)}]{scnews3}
\bibinfo{author}{\bibfnamefont{C.}~\bibnamefont{Huang}},
  \emph{\bibinfo{title}{Facebook and Twitter key to Arab Spring uprisings:
  report}} (\bibinfo{publisher}{The National}, \bibinfo{year}{June 6, 2011}), \urlprefix
  \href{http://www.thenational.ae/news/uae-news/facebook-and-twitter-key-to-arab-spring-uprisings-report}{\tt{http://www.thenational.ae/news/uae-news\linebreak /facebook-and-twitter-key-to-arab-spring-\linebreak uprisings-report}}.

\bibitem[{\citenamefont{Howard et~al.}(2011)\citenamefont{Howard, Duffy,
  Freelon, Hussain, Mari, and Mazaid}}]{scnews4}
\bibinfo{author}{\bibfnamefont{P.~N.} \bibnamefont{Howard}},
  \bibinfo{author}{\bibfnamefont{A.}~\bibnamefont{Duffy}},
  \bibinfo{author}{\bibfnamefont{D.}~\bibnamefont{Freelon}},
  \bibinfo{author}{\bibfnamefont{M.}~\bibnamefont{Hussain}},
  \bibinfo{author}{\bibfnamefont{W.}~\bibnamefont{Mari}}, \bibnamefont{and}
  \bibinfo{author}{\bibfnamefont{M.}~\bibnamefont{Mazaid}},
  \emph{\bibinfo{title}{Opening Closed Regimes: What Was the Role of Social
  Media During the Arab Spring?}} (\bibinfo{publisher}{Seattle: PIPTI},
  \bibinfo{year}{2011}), \urlprefix
 \href{http://pitpi.org/index.php/2011/09/11/opening-closed-regimes-what-was-the-role-of-social-media-during-the-arab-spring/}{\tt{http://pitpi.org/index.php/2011/09/11/\linebreak opening-closed-regimes-what-was-the-role-of\linebreak -social-media-during-the-arab-spring/}}.

\bibitem[{\citenamefont{Ball}(2012)}]{pb1cs}
\bibinfo{author}{\bibfnamefont{P.}~\bibnamefont{Ball}},
  \emph{\bibinfo{title}{Why Society is a Complex Matter}}
  (\bibinfo{publisher}{Springer}, \bibinfo{year}{2012}).

\bibitem[{\citenamefont{Castellano et~al.}(2009)\citenamefont{Castellano,
  Fortunato, and Loreto}}]{stat-phy-sc-1}
\bibinfo{author}{\bibfnamefont{C.}~\bibnamefont{Castellano}},
  \bibinfo{author}{\bibfnamefont{S.}~\bibnamefont{Fortunato}},
  \bibnamefont{and} \bibinfo{author}{\bibfnamefont{V.}~\bibnamefont{Loreto}},
  \bibinfo{journal}{Rev. Mod. Phys.} \textbf{\bibinfo{volume}{81}},
  \bibinfo{pages}{591} (\bibinfo{year}{2009}).

\bibitem[{\citenamefont{Onnela and Reed-Tsochas}(2010)}]{jp-on-1}
\bibinfo{author}{\bibfnamefont{J.}~\bibnamefont{Onnela}} \bibnamefont{and}
  \bibinfo{author}{\bibfnamefont{F.}~\bibnamefont{Reed-Tsochas}},
  \bibinfo{journal}{PNAS} \textbf{\bibinfo{volume}{107}},
  \bibinfo{pages}{18375} (\bibinfo{year}{2010}).

\bibitem[{\citenamefont{Centola}(2010)}]{centola1}
\bibinfo{author}{\bibfnamefont{D.}~\bibnamefont{Centola}},
  \bibinfo{journal}{Science} \textbf{\bibinfo{volume}{329}},
  \bibinfo{pages}{1194} (\bibinfo{year}{2010}).

\bibitem[{\citenamefont{Kempe et~al.}(2013)\citenamefont{Kempe, Kleinberg,
  Oren, and Slivkins}}]{dk1}
\bibinfo{author}{\bibfnamefont{D.}~\bibnamefont{Kempe}},
  \bibinfo{author}{\bibfnamefont{J.}~\bibnamefont{Kleinberg}},
  \bibinfo{author}{\bibfnamefont{S.}~\bibnamefont{Oren}}, \bibnamefont{and}
  \bibinfo{author}{\bibfnamefont{A.}~\bibnamefont{Slivkins}}, in
  \emph{\bibinfo{booktitle}{Proc. of EC}} (\bibinfo{year}{2013}).

\bibitem[{\citenamefont{Bond et~al.}(2012)\citenamefont{Bond, Fariss, Jones,
  Kramer, Marlow, Settle, and Fowler}}]{jp-on-2}
\bibinfo{author}{\bibfnamefont{R.~M.} \bibnamefont{Bond}},
  \bibinfo{author}{\bibfnamefont{C.~J.} \bibnamefont{Fariss}},
  \bibinfo{author}{\bibfnamefont{J.~J.} \bibnamefont{Jones}},
  \bibinfo{author}{\bibfnamefont{A.~D.~I.} \bibnamefont{Kramer}},
  \bibinfo{author}{\bibfnamefont{C.}~\bibnamefont{Marlow}},
  \bibinfo{author}{\bibfnamefont{J.~E.} \bibnamefont{Settle}},
  \bibnamefont{and} \bibinfo{author}{\bibfnamefont{J.~H.}
  \bibnamefont{Fowler}}, \bibinfo{journal}{Nature}
  \textbf{\bibinfo{volume}{489}}, \bibinfo{pages}{295} (\bibinfo{year}{2012}).

\bibitem[{\citenamefont{Moreno et~al.}(2004)\citenamefont{Moreno, Nekovee, and
  Pacheco}}]{rmr-1}
\bibinfo{author}{\bibfnamefont{Y.}~\bibnamefont{Moreno}},
  \bibinfo{author}{\bibfnamefont{M.}~\bibnamefont{Nekovee}}, \bibnamefont{and}
  \bibinfo{author}{\bibfnamefont{A.~F.} \bibnamefont{Pacheco}},
  \bibinfo{journal}{Phys. Rev. E} \textbf{\bibinfo{volume}{69}},
  \bibinfo{pages}{066130} (\bibinfo{year}{2004}).

\bibitem[{\citenamefont{Wasserman and Faust}(1994)}]{csnbook_2}
\bibinfo{author}{\bibfnamefont{S.}~\bibnamefont{Wasserman}} \bibnamefont{and}
  \bibinfo{author}{\bibfnamefont{K.}~\bibnamefont{Faust}},
  \emph{\bibinfo{title}{Social Network Analysis: Methods and Applications}}
  (\bibinfo{publisher}{Cambridge University Press}, \bibinfo{year}{1994}).

\bibitem[{\citenamefont{Vega-Redondo}(2007)}]{csnbook_1}
\bibinfo{author}{\bibfnamefont{F.}~\bibnamefont{Vega-Redondo}},
  \emph{\bibinfo{title}{Complex social networks}}
  (\bibinfo{publisher}{Cambridge University Press}, \bibinfo{year}{2007}).

\bibitem[{\citenamefont{Sood and Redner}(2005)}]{redner-3}
\bibinfo{author}{\bibfnamefont{V.}~\bibnamefont{Sood}} \bibnamefont{and}
  \bibinfo{author}{\bibfnamefont{S.}~\bibnamefont{Redner}},
  \bibinfo{journal}{Phys. Rev. Lett.} \textbf{\bibinfo{volume}{94}},
  \bibinfo{pages}{178701} (\bibinfo{year}{2005}).

\bibitem[{\citenamefont{Durrett et~al.}(2012)\citenamefont{Durrett, Gleeson,
  Lloyd, Mucha, Shi, Sivakoff, Socolar, and Varghes}}]{bill_2011}
\bibinfo{author}{\bibfnamefont{R.}~\bibnamefont{Durrett}},
  \bibinfo{author}{\bibfnamefont{J.~P.} \bibnamefont{Gleeson}},
  \bibinfo{author}{\bibfnamefont{A.~L.} \bibnamefont{Lloyd}},
  \bibinfo{author}{\bibfnamefont{P.~J.} \bibnamefont{Mucha}},
  \bibinfo{author}{\bibfnamefont{F.}~\bibnamefont{Shi}},
  \bibinfo{author}{\bibfnamefont{D.}~\bibnamefont{Sivakoff}},
  \bibinfo{author}{\bibfnamefont{J.~E.~S.} \bibnamefont{Socolar}},
  \bibnamefont{and} \bibinfo{author}{\bibfnamefont{C.}~\bibnamefont{Varghes}},
  \bibinfo{journal}{Proc. Natl. Acad. Sci.} \textbf{\bibinfo{volume}{109}}
  (\bibinfo{year}{2012}).

\bibitem[{\citenamefont{Vazquez et~al.}(2008)\citenamefont{Vazquez, Egu\'iluz,
  and Miguel}}]{vazquez_2008}
\bibinfo{author}{\bibfnamefont{F.}~\bibnamefont{Vazquez}},
  \bibinfo{author}{\bibfnamefont{V.}~\bibnamefont{Egu\'iluz}},
  \bibnamefont{and} \bibinfo{author}{\bibfnamefont{M.~S.}
  \bibnamefont{Miguel}}, \bibinfo{journal}{Phys. Rev. Letters.}
  \textbf{\bibinfo{volume}{100}}, \bibinfo{pages}{108702}
  (\bibinfo{year}{2008}).

\bibitem[{\citenamefont{Kimura and Hayakawa}(2008)}]{kimura_2008}
\bibinfo{author}{\bibfnamefont{D.}~\bibnamefont{Kimura}} \bibnamefont{and}
  \bibinfo{author}{\bibfnamefont{Y.}~\bibnamefont{Hayakawa}},
  \bibinfo{journal}{Phys. Rev. E} \textbf{\bibinfo{volume}{78}},
  \bibinfo{pages}{016103} (\bibinfo{year}{2008}).

\bibitem[{\citenamefont{Weidlich}(2000)}]{ww1}
\bibinfo{author}{\bibfnamefont{W.}~\bibnamefont{Weidlich}},
  \emph{\bibinfo{title}{Sociodynamics : A Systematic Approach to Modelling the
  Social Sciences}} (\bibinfo{publisher}{Harwood, Academic Amsterdam},
  \bibinfo{year}{2000}).

\bibitem[{\citenamefont{Schelling}(1978)}]{tc1}
\bibinfo{author}{\bibfnamefont{T.~C.} \bibnamefont{Schelling}},
  \emph{\bibinfo{title}{Micromotives and Macrobehaviour}}
  (\bibinfo{publisher}{W.W. Norton, New York}, \bibinfo{year}{1978}).

\bibitem[{\citenamefont{Timpanaro and Prado}(2009)}]{szm1}
\bibinfo{author}{\bibfnamefont{A.~M.} \bibnamefont{Timpanaro}}
  \bibnamefont{and} \bibinfo{author}{\bibfnamefont{C.~P.~C.}
  \bibnamefont{Prado}}, \bibinfo{journal}{Phys. Rev. E}
  \textbf{\bibinfo{volume}{80}}, \bibinfo{pages}{021119}
  (\bibinfo{year}{2009}).

\bibitem[{\citenamefont{Zschaler et~al.}(2012)\citenamefont{Zschaler,
  B\"{o}hme, Sei?inger, Huepe, and Gross}}]{thilo1}
\bibinfo{author}{\bibfnamefont{G.}~\bibnamefont{Zschaler}},
  \bibinfo{author}{\bibfnamefont{G.~A.} \bibnamefont{B\"{o}hme}},
  \bibinfo{author}{\bibfnamefont{M.}~\bibnamefont{Sei?inger}},
  \bibinfo{author}{\bibfnamefont{C.}~\bibnamefont{Huepe}}, \bibnamefont{and}
  \bibinfo{author}{\bibfnamefont{T.}~\bibnamefont{Gross}},
  \bibinfo{journal}{Phys. Rev. E} \textbf{\bibinfo{volume}{85}},
  \bibinfo{pages}{046107} (\bibinfo{year}{2012}).

\bibitem[{\citenamefont{B\"{o}hme and Gross}(2012)}]{thilo2}
\bibinfo{author}{\bibfnamefont{G.~A.} \bibnamefont{B\"{o}hme}}
  \bibnamefont{and} \bibinfo{author}{\bibfnamefont{T.}~\bibnamefont{Gross}},
  \bibinfo{journal}{Phys. Rev. E} \textbf{\bibinfo{volume}{85}},
  \bibinfo{pages}{066117} (\bibinfo{year}{2012}).

\bibitem[{\citenamefont{Gleeson et~al.}(2013)\citenamefont{Gleeson, Cellai,
  Onnela, Porter, and Reed-Tsochas}}]{gleeson1}
\bibinfo{author}{\bibfnamefont{J.~P.} \bibnamefont{Gleeson}},
  \bibinfo{author}{\bibfnamefont{D.}~\bibnamefont{Cellai}},
  \bibinfo{author}{\bibfnamefont{J.-P.} \bibnamefont{Onnela}},
  \bibinfo{author}{\bibfnamefont{M.~A.} \bibnamefont{Porter}},
  \bibnamefont{and}
  \bibinfo{author}{\bibfnamefont{F.}~\bibnamefont{Reed-Tsochas}}
  (\bibinfo{year}{2013}), \eprint{arXiv:1305.7440}.

\bibitem[{\citenamefont{Redner}(1998)}]{redner-1}
\bibinfo{author}{\bibfnamefont{S.}~\bibnamefont{Redner}},
  \bibinfo{journal}{Europ. Phys. J. B} \textbf{\bibinfo{volume}{4}},
  \bibinfo{pages}{131} (\bibinfo{year}{1998}).

\bibitem[{\citenamefont{Volovik and Redner}(2012)}]{redner-2}
\bibinfo{author}{\bibfnamefont{D.}~\bibnamefont{Volovik}} \bibnamefont{and}
  \bibinfo{author}{\bibfnamefont{S.}~\bibnamefont{Redner}},
  \bibinfo{journal}{J. Stat. Mech.} \textbf{\bibinfo{volume}{P04003}}
  (\bibinfo{year}{2012}).

\bibitem[{\citenamefont{Barnett and Casper}(2001)}]{se1}
\bibinfo{author}{\bibfnamefont{E.}~\bibnamefont{Barnett}} \bibnamefont{and}
  \bibinfo{author}{\bibfnamefont{M.}~\bibnamefont{Casper}},
  \bibinfo{journal}{American Journal of Public Health}
  \textbf{\bibinfo{volume}{91}} (\bibinfo{year}{2001}).

\bibitem[{\citenamefont{Herman and Chomsky}(1988)}]{nc1}
\bibinfo{author}{\bibfnamefont{E.~S.} \bibnamefont{Herman}} \bibnamefont{and}
  \bibinfo{author}{\bibfnamefont{N.}~\bibnamefont{Chomsky}},
  \emph{\bibinfo{title}{Manufacturing Consent: The Political Economy of the
  Mass Media}} (\bibinfo{publisher}{Pantheon Books}, \bibinfo{year}{1988}).

\bibitem[{\citenamefont{Watts and Strogatz}(1998)}]{strogatz_1998}
\bibinfo{author}{\bibfnamefont{D.~J.} \bibnamefont{Watts}} \bibnamefont{and}
  \bibinfo{author}{\bibfnamefont{S.}~\bibnamefont{Strogatz}},
  \bibinfo{journal}{Nature} \textbf{\bibinfo{volume}{393}},
  \bibinfo{pages}{440} (\bibinfo{year}{1998}).

\bibitem[{\citenamefont{McPherson et~al.}(2001)\citenamefont{McPherson,
  Smith-Lovin, and Cook}}]{homo1}
\bibinfo{author}{\bibfnamefont{M.}~\bibnamefont{McPherson}},
  \bibinfo{author}{\bibfnamefont{L.}~\bibnamefont{Smith-Lovin}},
  \bibnamefont{and} \bibinfo{author}{\bibfnamefont{J.~M.} \bibnamefont{Cook}},
  \bibinfo{journal}{Annu. Rev. Sociol.} \textbf{\bibinfo{volume}{27}},
  \bibinfo{pages}{415} (\bibinfo{year}{2001}).

\bibitem[{\citenamefont{Filho et~al.}(1999)\citenamefont{Filho, Almeida, Jr.,
  and Moreira}}]{rncost1}
\bibinfo{author}{\bibfnamefont{R.~N.~C.} \bibnamefont{Filho}},
  \bibinfo{author}{\bibfnamefont{M.~P.} \bibnamefont{Almeida}},
  \bibinfo{author}{\bibfnamefont{J.~S.~A.} \bibnamefont{Jr.}},
  \bibnamefont{and} \bibinfo{author}{\bibfnamefont{J.~E.}
  \bibnamefont{Moreira}}, \bibinfo{journal}{Phys. Rev. E}
  \textbf{\bibinfo{volume}{60}}, \bibinfo{pages}{1067} (\bibinfo{year}{1999}).

\bibitem[{\citenamefont{Maulana and Situngkir}(2011)}]{els1}
\bibinfo{author}{\bibfnamefont{A.}~\bibnamefont{Maulana}} \bibnamefont{and}
  \bibinfo{author}{\bibfnamefont{H.}~\bibnamefont{Situngkir}},
  \emph{\bibinfo{title}{Power laws in Elections A Survey}}
  (\bibinfo{publisher}{Bandung Fe Institute}, \bibinfo{year}{2011}), \urlprefix 
  \href{http://cogprints.org/6934/}{\tt{http://cogprints.org/6934/}}.

\bibitem[{\citenamefont{Farmer and Geanakoplos}(2006)}]{frd1}
\bibinfo{author}{\bibfnamefont{J.}~\bibnamefont{Farmer}} \bibnamefont{and}
  \bibinfo{author}{\bibfnamefont{J.}~\bibnamefont{Geanakoplos}},
  \emph{\bibinfo{title}{Power laws in economics and elsewhere}}
  (\bibinfo{publisher}{Tech. Rep., Santa Fe Institute}, \bibinfo{year}{2006}).

\bibitem[{\citenamefont{Clauset et~al.}(2009)\citenamefont{Clauset, Shalizi,
  and Newman}}]{accr1}
\bibinfo{author}{\bibfnamefont{A.}~\bibnamefont{Clauset}},
  \bibinfo{author}{\bibfnamefont{C.~R.} \bibnamefont{Shalizi}},
  \bibnamefont{and} \bibinfo{author}{\bibfnamefont{M.~E.~J.}
  \bibnamefont{Newman}}, \bibinfo{journal}{SIAM Review}
  \textbf{\bibinfo{volume}{51 (4)}}, \bibinfo{pages}{661}
  (\bibinfo{year}{2009}).

\bibitem[{\citenamefont{Stumpf and Porter}(2012)}]{mph1}
\bibinfo{author}{\bibfnamefont{M.}~\bibnamefont{Stumpf}} \bibnamefont{and}
  \bibinfo{author}{\bibfnamefont{M.}~\bibnamefont{Porter}},
  \bibinfo{journal}{Science} \textbf{\bibinfo{volume}{335}},
  \bibinfo{pages}{665} (\bibinfo{year}{2012}).

\bibitem[{\citenamefont{Douglas et~al.}(2011)\citenamefont{Douglas, Pra??t,
  and Zhangb}}]{ems1}
\bibinfo{author}{\bibfnamefont{A.}~\bibnamefont{Douglas}},
  \bibinfo{author}{\bibfnamefont{H.~P.} \bibnamefont{Pra??t}},
  \bibnamefont{and} \bibinfo{author}{\bibfnamefont{C.-Q.}
  \bibnamefont{Zhangb}}, \bibinfo{journal}{Proc. Natl. Acad. Sci.}
  \textbf{\bibinfo{volume}{108}}, \bibinfo{pages}{8605} (\bibinfo{year}{2011}).

\bibitem[{\citenamefont{Fossett}(2006)}]{ems2}
\bibinfo{author}{\bibfnamefont{M.}~\bibnamefont{Fossett}}, \bibinfo{journal}{J
  Math Sociol.} \textbf{\bibinfo{volume}{30}}, \bibinfo{pages}{185}
  (\bibinfo{year}{2006}).

\bibitem[{\citenamefont{Pollicott and Weiss}(2001)}]{ems3}
\bibinfo{author}{\bibfnamefont{M.}~\bibnamefont{Pollicott}} \bibnamefont{and}
  \bibinfo{author}{\bibfnamefont{H.}~\bibnamefont{Weiss}},
  \bibinfo{journal}{Adv Appl Math.} pp. \bibinfo{pages}{17--40}
  (\bibinfo{year}{2001}).

\bibitem[{\citenamefont{Dorogovtsev}(2010)}]{sndoro1}
\bibinfo{author}{\bibfnamefont{S.~N.} \bibnamefont{Dorogovtsev}},
  \emph{\bibinfo{title}{Lectures on Complex Networks}}
  (\bibinfo{publisher}{Oxford University Press,Oxford}, \bibinfo{year}{2010}).

\bibitem[{\citenamefont{Newman}(2010)}]{nw1}
\bibinfo{author}{\bibfnamefont{M.~E.~J.} \bibnamefont{Newman}},
  \emph{\bibinfo{title}{Networks: An Introduction}} (\bibinfo{publisher}{Oxford
  University Press,Oxford}, \bibinfo{year}{2010}).

\bibitem[{\citenamefont{Toral and Tessone}(2007)}]{fts1}
\bibinfo{author}{\bibfnamefont{R.}~\bibnamefont{Toral}} \bibnamefont{and}
  \bibinfo{author}{\bibfnamefont{C.~J.} \bibnamefont{Tessone}},
  \bibinfo{journal}{Commun. Comput. Phys.} \textbf{\bibinfo{volume}{2}},
  \bibinfo{pages}{177} (\bibinfo{year}{2007}).

\bibitem[{\citenamefont{Jones et~al.}(2001--)\citenamefont{Jones, Oliphant,
  Peterson et~al.}}]{scipy-1}
\bibinfo{author}{\bibfnamefont{E.}~\bibnamefont{Jones}},
  \bibinfo{author}{\bibfnamefont{T.}~\bibnamefont{Oliphant}},
  \bibinfo{author}{\bibfnamefont{P.}~\bibnamefont{Peterson}},
  \bibnamefont{et~al.}, \emph{\bibinfo{title}{{SciPy}: Open source scientific
  tools for {Python}}} (\bibinfo{year}{2001--}),
  \urlprefix \href{http://www.scipy.org/}{\tt{http://www.scipy.org/}}.

\bibitem[{\citenamefont{Wasserman}(1994)}]{snet1}
\bibinfo{author}{\bibfnamefont{S.}~\bibnamefont{Wasserman}},
  \emph{\bibinfo{title}{Social network analysis : methods and applications}}
  (\bibinfo{publisher}{Cambridge University Press, Cambridge},
  \bibinfo{year}{1994}).

\bibitem[{\citenamefont{Holland and Leinhardt}(1971)}]{trans_1971}
\bibinfo{author}{\bibfnamefont{P.~W.} \bibnamefont{Holland}} \bibnamefont{and}
  \bibinfo{author}{\bibfnamefont{S.}~\bibnamefont{Leinhardt}},
  \bibinfo{journal}{Comparative Group Studies} \textbf{\bibinfo{volume}{2}},
  \bibinfo{pages}{107} (\bibinfo{year}{1971}).

\bibitem[{\citenamefont{Schofield and Sened}(2006)}]{mdemo-1}
\bibinfo{author}{\bibfnamefont{N.}~\bibnamefont{Schofield}} \bibnamefont{and}
  \bibinfo{author}{\bibfnamefont{I.}~\bibnamefont{Sened}},
  \emph{\bibinfo{title}{Multiparty Democracy: Elections and Legislative
  Politics}} (\bibinfo{publisher}{Cambridge University Press},
  \bibinfo{year}{2006}).

\bibitem[{\citenamefont{Webster}(2003)}]{mdemo-2}
\bibinfo{author}{\bibfnamefont{T.~J.} \bibnamefont{Webster}},
  \emph{\bibinfo{title}{Managerial Economics : Theory and Practices}}
  (\bibinfo{publisher}{Academic Press}, \bibinfo{year}{2003}).

\end{thebibliography}

\end{document}